\documentclass[a4paper,10pt]{article}   

  \usepackage{makeidx}
  \usepackage{graphicx,graphics,color}
  \usepackage{latexsym,doc,fontenc,exscale}
\usepackage{amsmath,amsfonts,amsthm,eucal,amssymb}

\usepackage{multicol}
\usepackage{times}
\usepackage{pstricks}

 \newcommand{\sn}{\textrm{sn}}
 \newcommand{\cn}{\textrm{cn}}
 
 \newcommand{\nss}{\textrm{ns}}

 \newcommand{\dc}{\textrm{dc}}
 \newcommand{\cd}{\textrm{cd}}
 \newcommand{\ds}{\textrm{ds}}
 \newcommand{\sd}{\textrm{sd}}
 \newcommand{\nd}{\textrm{nd}}
 \newcommand{\ws}{\textrm{c}}

\begin{document}
 \title{\bf New Exact Solutions of a Generalized Shallow Water Wave
 Equation}\vspace{2cm}

\author{\vspace{1cm}\Large \textbf{Bijan Bagchi}$^{1}$\footnote{bbagchi123@rediffmail.com },
\textbf{Supratim Das}$^{1}$\footnote{supratimiitkgp@gmail.com} and \textbf{Asish Ganguly}$^{2}$\footnote{aganguly@maths.iitkgp.ernet.in, gangulyasish@rediffmail.com}\\
$^{1}$Department of Applied Mathematics, University of Calcutta,\\ 92 Acharya Prafulla Chandra Road, Kolkata-700009, India\\
 \vspace{0.1cm}$^{2}$ Department of Mathematics, Indian Institute of Technology, Kharagpur 721302, India}
 \huge{\date{February 16, 2010}}
\maketitle


\vspace{2cm} \noindent
 \begin{abstract}
 \large
 In this work an extended elliptic function method
 is proposed and applied to the generalized shallow water wave equation.
  We systematically investigate to classify new exact travelling wave
 solutions expressible in terms of quasi-periodic elliptic integral function and doubly-periodic Jacobian
 elliptic functions. The derived new solutions include rational, periodic, singular
 and solitary wave solutions. An interesting comparison with the
 canonical procedure is provided. In some cases the obtained elliptic solution has singularity
 at certain region in the whole space. For such solutions we have computed the
 effective region where the obtained solution is free from such a
 singularity.
 \end{abstract}

 \vspace{0.8cm}
 \noindent

\large{Keywords: \textbf{\it Shallow Water Wave Equation,
Integrable Systems, Travelling Waves,\\ \hspace*{2.3cm} Jacobian
Elliptic Function, Rational Solution, Singular Solution.}}

\vspace{0.8cm}
 \noindent
 PACS number(s): \textbf{\it 02.30.Jr, 02.30.Ik, 05.45.-a}
\twocolumn \nopagebreak[4]
 \section{Introduction}

Seeking analytical as well as numerical solutions of nonlinear
systems has continued to attract attention through the last few
decades \cite{lev,kon,leo,nay,her}. Mathematically, nonlinear
equations do not normally have solutions which would superpose
making the systems they represent rather complicated and difficult
to analyze \cite{wig,guh}. On the other hand, an extensive study
of a number of nonlinear systems has revealed that there do exist
solutions which are not only interesting in their own right but
have a wide range of applicability \cite{mob,mak}. To generate
exact solutions and to understand their properties several
important techniques have been developed such as the inverse
scattering approach \cite{zak}, Lax pair formulation \cite{lax},
Backlund transformations \cite{wanl} and Hirota's bilinear method
\cite{hirota1,hirota2}. Nonlinear equations have also been shown
to arise from a boundary value problem \cite{whi} possessing
hierarchy of conservation laws \cite{miu}.

An important class of solutions of nonlinear evolution equations
is concerned with those of the travelling waves that  reduce the
guiding partial differential equation of two variables namely, $x$
and $t$ to an ordinary differential equation of one independent
variable $z=x-\ws t$ where $\ws(\in\mathbb{R}-\{0\})$ is a
parameter signifying the speed with which the wave travels either
to the right or left. A number of methods has been employed in the
literature to obtain the travelling wave solutions of various
types. Of these the tanh-method stands out as one of the very
effective tools for solving certain classes of nonlinear evolution
and wave equation \cite{mal}. Others include the variational
iteration method for a class of linear and nonlinear equations
\cite{waz1}, homogeneous balance method \cite{wan}, the hyperbolic
method \cite{waz}, the trigonometric method \cite{yan}, Darboux's
transformation \cite{wan3}, F-function method \cite{yom},
$G'/G$-expansion method \cite{wang}, a unified algebraic method
\cite{fan} and more.

Recently the existence of travelling wave solutions for the
generalized shallow water wave (GSWW) equation \cite{hie}
\begin{equation}\label{GSWW}
  u_{xxxt}+\alpha u_{x}u_{xt}+\beta
 u_{t}u_{xx}-u_{xt}-u_{xx}=0\, , \:\: (u_l\equiv \frac{\partial u}{\partial l})
 \end{equation}
 with $\alpha,~\beta\in\mathbb{R}-\{0\}$
has been noted. The derivation of (\ref{GSWW}) follows from the
classical water wave theory with the aid of Boussinesq
approximation. In an interesting review, Clarkson and Mansfield
\cite{cla} considered the various classical and non-classical
reductions of the GSWW equation wherein they also investigated the
Painlev\'{e} tests to examine the complete integrability of
(\ref{GSWW}) which holds if and only if $\alpha=\beta$ or
$\alpha=2\beta$. They explored some simple and non-trivial family
of solutions of (\ref{GSWW}) while in \cite{elw} a class of exact
travelling wave solutions were obtained by making use of the
homogeneous balance method and a modified hyperbolic method.

In this article we plan to study the travelling wave solutions
\begin{equation}\label{travelling-wave}
 u(x,t)\equiv u(z)\, ,\quad z= x-\ws t\, ,
 \end{equation}
of the GSWW equation (\ref{GSWW}) in a more general framework. We
show that the canonical procedure namely, the classical technique
\cite{dra}, based on an integration within a suitable range not
only recovers both known periodic (including elliptic and
trigonometric) and non-periodic hyperbolic solutions, but also
unfolds several new elliptic and rational solutions. However such
a constructive method has its limitations in the sense that the
range of validity of the  obtained solutions is fixed by the
analysis of the zeros of the governing cubic polynomial in $u'(z)$
beforehand and so, in practical applications, a particular
solution proves difficult to implement.

One of our aims is to go beyond this standard path by proposing an
extended method based on the use of Jacobian elliptic functions.
We refer to it as an extended elliptic function (EEF) method. In
this context the works of Refs \cite{yom2} and \cite{che} worth to
be mentioned which used the elliptic functions for a class of
nonlinear evolution equations. These works consider a finite
series expansion in the form
\begin{equation}\label{pos-power}
v(z)\equiv u'(z)=\sum_{i=0}^{m} a_{i}F^{i}(z)\,,
\end{equation}
 to generate travelling wave solutions. In (\ref{pos-power}) the positive
 integer $m$ is determined by balancing the coefficients $a_i$'s after
 substituting the expansion in the given nonlinear equation and comparing
 the relevant powers. Our EEF method
 generalizes this approach by including the negative powers of $F$ thereby
 addressing the full-range series as given in the following
\begin{equation}\label{finite-series}
   v(z)=\sum_{i=-m}^{m} a_{i}F^{i}(z)\,.
\end{equation}
In Equation~(\ref{finite-series}), F is an unknown functional
constrained to satisfy the relation
\begin{equation}\label{constraint}
    (F')^{2}=(1+\varepsilon_1 F^{2})(\varepsilon_2+\varepsilon_3F^{2})\, ,
\end{equation}
where the prime denotes the derivative with respect to $z$
throughout the text. The important feature of the EEF method is
that it includes not only the positive integral powers of $F$ but
also the negative ones. Our motivation for keeping lower negative
degrees of F comes from the fact that in typical solvable systems
like the Korteweg de Vries the related spectral problem allows
similar extensions \cite{and1,and2}. It has been shown that these
lower equations may arise \cite{and2} as a necessary part of an
extended symmetry structure \cite{bag1} while it is known
\cite{bag2} that the underlying commutator generates the whole set
of symmetry operators for the system.

It is not difficult to see that the biquadratic integral
\cite{as4} emerging from the constraint (\ref{constraint})
facilitates expressing the travelling wave solutions in terms of
the Jacobian elliptic functions $\textrm{pq} (z-z_0,k)$, where the
symbol pq represents twelve such distinct functions
sn,cn,dn,ns($\equiv 1/$sn),cs($\equiv$ cn/sn) etc. The novelty of
the EEF method is that it opens up a broad spectra of new
travelling wave solutions that include the already known ones (see
for example, \cite{dra,che,kha,fuz,yus}). We shall explicitly
provide the solutions generated from the EEF method for some
special selections of the integral parameters.

  This article is organized as follows. In Section~\ref{classical}, we derive
  the reduced ordinary differential equation for the GSWW equation and then go
  for the classical technique to obtain exact analytical solutions. In Section~\ref{new-method}
  we turn to the constraint (\ref{constraint}) to take up the construction of our new procedure and then categorise
  different types of solutions being generated from the EEF method by appealing to
  particular choices of the
  integral parameters. It will be seen that the solutions in the  appropriate
  limits reduce to previously known results.  Finally, in Section~\ref{conclusion} we present
  a summary of our results.

 \section{Canonical procedure for the \\ GSWW equation}\label{classical}

To start with, we substitute the form (\ref{travelling-wave}) into
the GSWW equation (\ref{GSWW}) that results in a fourth order
ordinary differential equation
\begin{equation}\label{ODE}
\hspace{-1cm}\ws u''''+(\alpha+\beta)\ws u'u''+(1-\ws)u''=0\, .
\end{equation}
  Let us note that the parameters $\alpha,\beta$ controls the strength of non-linearity
  of above equation. Without loss of generality, one can leave one of them, say $\alpha$,
  arbitrary while the other parameter $\beta$ may be varied to tune the strength of non-linearity
   appropriate for the concerned physical situations. Three particular values of $\beta$ namely
   $\beta=\pm \alpha,\alpha/2$ are known to correspond completely integrable system. In this
   article we shall explore the solutions of Equation~(\ref{GSWW}) leaving both of them arbitrary
   except for $\beta=-\alpha$ as it points to the linear system and hence is well-known. It is straightforward
  to obtain a second integral of (\ref{ODE}) given by
\begin{equation}\label{2nd-integral}
  3\ws u''^2+\ws (\alpha+\beta)u'^3+3(1-\ws)u'^2+6\ws c_1u'+3\ws c_2=0\, ,
\end{equation}
which may be expressed as a first order equation in terms of the
variable $v(z)\equiv u'(z)$:
\begin{equation}\label{1st-order}
   v'^2=-\frac{\alpha+\beta}{3}\mathcal{P}_3(v)\equiv
    -\frac{\alpha+\beta}{3}\prod_{j=1}^3(v-v_j)\, .
\end{equation}
  Note that in Equation~(\ref{2nd-integral}) and also elsewhere in the text, the arbitrary
  constants appearing through the process of integration are denoted  by the
  symbol $c_j$ and will not be explicitly mentioned further. In
  (\ref{1st-order})
  the monic cubic polynomial $\mathcal{P}_3(v)$ is given by
 \begin{equation}\label{cubic}
   \hspace{-1cm} \mathcal{P}_3(v)=v^{3}+\frac{3(1-\ws)}{\ws (\alpha+\beta)}v^{2}+
 \frac{6c_{1}}{\alpha+\beta}v+\frac{3c_{2}}{\alpha+\beta}\, .
\end{equation}
 For simplicity let us assume that all the three roots of $\mathcal{P}_3(v)$ are real
 and focus on the arrangement $v_1\geq v_2\geq v_3$. We consider first the non-degenerate case
 of distinct roots for $\alpha+\beta<0$. A formal integration of (\ref{1st-order}) in the
 range $[(v_j,z_j),(v,z)]$, where the point $z_j$ is to be determined from the transcendental
 equation $v(z_j)=v_j$, yields
\begin{equation}\label{integration}
   \hspace{-1cm} \int_{z_{j}}^{z}\textrm{d} z=\pm\int_{v_{j}}^{v}\frac{\textrm{d} w}{[(-\frac{\alpha+\beta}{3})\prod_{j=1}^3 (w-v_j)]^{1/2}}\, .
\end{equation}
  Now care is to be taken to choose the root $v_j, j=1,2 \mbox{ or }3$ that defines the
  interval of validity of the solution. It is easy to see  that the reality condition for $v(z)$ gives
  rise to two different cases namely either $v_3<v<v_2$ or $v_1<v<\infty$. We address them by turn below.

\vspace*{0.5cm} \noindent
 {\large \bf Case 1:$\qquad \mathbf{v_3<v<v_2 \quad (\boldsymbol{\alpha+\beta}<0)}$}

\vspace*{0.2cm}

Keeping in mind that the range of integration is between $[v_3,v]$
 and $[z_3,z]$, the following change of variables
$w\rightarrow \varphi$ proves useful
\begin{equation}\label{substitution-1}
    w=v_{3}+(v_{2}-v_{3})\sin^{2}\varphi\, .
\end{equation}
The standard definition for Jacobian elliptic function then
expresses $v(z)$ in a closed analytic form
\begin{equation}\label{sol-v-1}
    v(z)\equiv
    u'(z)=v_{3}+(v_{2}-v_{3})\sn^{2}(\nu (z-z_{3}),k)\, ,
\end{equation}
 where $\nu=\surd [|\alpha+\beta|(v_1-v_3)/3]/2$ and
 $k^2=(v_2-v_3)/(v_1-v_3)(0<k^2<1)$ is the modulus of the
 elliptic function \cite{ste}. In terms of the elliptic
 integral function of the second kind $E(\varphi_{\ell},k)$ through the
 relation $\sin\varphi_{\ell}=\sn(\nu (z-z_{\ell}),k)$, defined by the
 integral
 \begin{equation}\label{elliptic-integral-func-2nd}
    E(\varphi_{\ell},k)=\int_{0}^{\varphi_{\ell}}\sqrt{1-k^2\sin^2\alpha}\: \: \textrm{d}\alpha\, , \qquad \ell=1,3\, ,
\end{equation}
 we obtain $u(z)$, after integrating (\ref{sol-v-1}), in the final form
\begin{equation}\label{final-sol-1}
u(z)=u_{3}+\frac{v_2-k'^2v_3}{k^2}(z-z_{3})-\frac{v_{2}-v_{3}}{\nu
k^2}E(\varphi_3,k)\, .
\end{equation}
 In the above solution and also in the following $u_j$ is defined in a natural way as $u_j=u(z_j), j=1,2,3$
 while $k'^2=1-k^2$ is the complementary modulus of
 elliptic functions.

 It may be pointed out that (\ref{final-sol-1}) is a new travelling wave solution. Two interesting
 limits  namely $k\rightarrow 1$ and $k\rightarrow 0$ may be of interest  to  notice that lead to the
 existence of a double root of $\mathcal{P}_3(v)$. For  $k\rightarrow 1\mbox{-\small{0}}$, we must
 let $v_1\rightarrow v_2\mbox{\small{+0}}$, but then $v_3$ cannot be allowed to coincide with  $v_2$ as this
 will imply that the limiting solution has no range of validity. Along the same reasoning we conclude that the
 limit $k\rightarrow 0$ is disallowed for this case. Noting that $E(\varphi_3,k\rightarrow 1)=\tanh [\nu(z-z_3)]$ the solution (\ref{final-sol-1})
 thus reduces to the  hyperbolic form for $k\rightarrow 1\mbox{-\small{0}}$
 \begin{equation}\label{hyperbolic-reduction-1}
    u(z)=u_3+v_2(z-z_3) -\frac{v_{2}-v_{3}}{\nu }\tanh [\nu (z-z_3)]\, .
\end{equation}

 \vspace*{0.5cm}
 \noindent
 {\large \bf Case 2:$\qquad \mathbf{v_1<v<\infty \quad (\boldsymbol{\alpha+\beta}<0)}$}

\vspace*{0.2cm}

Clearly the range of integration now lies between $[v_1,v]$ and
$[z_1,z]$. Employing the substitution
\begin{equation}\label{substitution-2}
    w=v_{3}+(v_{1}-v_{3})\textrm{cosec}^{2}\varphi\, ,
\end{equation}
we obtain the following expression for $v(z)$
\begin{equation}\label{sol-v-2}
 v(z)\equiv
    u'(z)=v_{3}+(v_{1}-v_{3})\dc^{2}(\nu (z-z_{1}),k)\, .
\end{equation}
The above equation, on further integration, yields another new
travelling wave solution
\begin{eqnarray}\label{final-sol-2}
 \hspace{-1cm} u(z)&=&u_{1}+v_{1}(z-z_{1})-\frac{v_{1}-v_{3}}{\nu} \left [ E(\varphi_1,k)\right . \nonumber\\
 \hspace{-1cm} &&\mbox{}\left .  -\sn (\nu(z-z_1),k)\dc (\nu(z-z_1),k)\right ] \, .
\end{eqnarray}
In above we have used the relation $\sn (z+K,k)=\cd (z,k)$, where
$K(k)=\int_0^{\pi/2}\textrm{d}\alpha/\sqrt{1-k^2\sin^2 \alpha}$ is
the quarter real period of elliptic sine function. Let us now take
the limit $v_2\rightarrow v_1\mbox{-\small{0}}$ keeping
 $v_3<v_2$ that makes the travelling wave solution linear in
 spatial and time coordinates
 \begin{equation}\label{linear-reduction-2}
  \hspace{-1cm}  k\rightarrow 1\mbox{-\small{0}}: u(x,t)=(u_1-v_1z_1)+v_1(x-\ws t)\, .
\end{equation}
On the other hand, letting $v_3\rightarrow v_2\mbox{-\small{0}}$
where $v_2<v_1$, we recover the periodic solution
\begin{equation}\label{periodic-reduction-2}
  k\rightarrow 0: u(z)=u_1+v_2(z-z_1)+\frac{v_1-v_2}{\nu}\tan [\nu(z-z_1)]\, .
\end{equation}

The case for $\alpha+\beta>0$ will give the result in
complementary frame which can be derived along the same way as
before. Without giving the details of the procedure, we are
providing below the final solutions.

\vspace*{0.5cm} \noindent
 {\large \bf Case 3:$\qquad \mathbf{v_2<v<v_1 \quad (\boldsymbol{\alpha+\beta}>0)}$}
\begin{eqnarray}\label{final-sol-3}
 \hspace{-0.5cm} v(z) &\hspace{-7pt}=\hspace{-7pt}& v_1-(v_1-v_2)\sn^{2}(\nu (z-z_1),k')\, ,\nonumber \\
 \hspace{-0.5cm} u(z) &\hspace{-7pt}=\hspace{-7pt}& u_1+\frac{v_2-k^2v_1}{k'^2}(z-z_1)+\frac{v_1-v_2}{\nu k'^2}E(\varphi'_1,k')\, ,
\end{eqnarray}
where $\sin\varphi'_i=\sn (\nu(z-z_1),k'),i=1,3$. To derive the
above we made the following substitution in (\ref{integration})
\begin{equation}\label{substitution-3}
     w=v_1+(v_2-v_1)\sin^2\varphi\, .
\end{equation}

\vspace*{0.5cm} \noindent
 {\large \bf Case 4:$\qquad \mathbf{-\infty<v<v_3 \quad (\boldsymbol{\alpha+\beta}>0)}$}
\begin{eqnarray}\label{final-sol-4}
 \hspace{-1cm} v(z) &=& v_1-(v_1-v_3)\dc^{2}(\nu (z-z_3),k')\, ,\nonumber \\
 \hspace{-1cm} u(z) &=& u_3+v_3(z-z_3)+\frac{v_1-v_3}{\nu}\left [ E(\varphi'_3,k')\right . \\
 \hspace{-1cm} && \mbox{}\left . -\sn (\nu(z-z_3),k')\dc (\nu(z-z_3),k')\right ]\, ,\nonumber
\end{eqnarray}
which comes from the following substitution in (\ref{integration})
\begin{equation}\label{substitution-4}
     w=v_1-(v_1-v_3)\textrm{cosec}^2\varphi\, .
\end{equation}
It is interesting to note that in contrast to the Cases \textbf{1}
and \textbf{2} for $\alpha+\beta<0$, the known hyperbolic and
periodic solutions are recovered for $\alpha+\beta>0$ (Cases
\textbf{3,4}) in the complementary limit of the elliptic solutions
(\ref{final-sol-3}) and (\ref{final-sol-4}).

Finally let us deal with the degenerate case when
$\mathcal{P}_3(v)$ has a triple zero, i.~\hspace{-3pt}e.\
$v_1=v_2=v_3=v_0$. From the cubic (\ref{cubic}) it is clear that
we have to choose $c_2=-2v_0c_1/3$ giving the multiple root
$v_0=(\ws-1)/[\ws(\alpha+\beta)]$. The range of integration will
be $-\infty<v<v_0$ or $v_0<v<\infty$ according as
$\alpha+\beta\gtrless0$. The singular rational solution can be
expressed in a compact form as follows
\begin{equation}\label{rational-triple}
   \mathcal{P}_3(v)=(v-v_0)^3: \quad  u=v_0(z-c_3)+\frac{12}{\alpha+\beta}\frac{1}{z-c_3}+c_4\, ,
\end{equation}
which was obtained in \cite{elw} from a different approach.

We have thus obtained several new solutions by applying the
classical technique based on an analysis of the zeros of a cubic
polynomial. Interestingly in the appropriate limits all the known
solutions are retrieved. It is clear that for practical
applications of the solutions $u(z)$ derived in this section, a
knowledge about the position of zeros of  $u''(z)$ is required
which somewhat weakens their utility. In the next section we
propose a new method that is free from such a limitation.
 \section{The EEF method for the GSWW equation}\label{new-method}

Here instead of focussing on the zeros of $u''(z)$ which are
unknown we fix the zeros of $u'(z)$. Since $u''$ has three zeros
[see Equation~(\ref{1st-order}) of Sec.~\ref{classical}], it is
obvious that $u'$ has four zeros. This motivates us to propose the
following construction.

\subsection{Construction}\label{construction-EEF}

Let us turn to the fourth-order ordinary differential equation
(\ref{ODE}) which we integrate to write
\begin{equation}\label{2nd-order}
  \hspace{-1cm}  2\ws v''+\ws (\alpha+\beta)v^{2}+2(1-\ws)v-c_{5}=0.
\end{equation}
We look for a formal solution in the form
\begin{equation}\label{proposed-expansion}
   v(z)=\sum_{i=-m}^{m}a_iF^i\, ,
\end{equation}
where the generating functional $F(z(x,t))$ satisfies the
constraint~(\ref{constraint}). Note that the inclusion of negative
powers of $F$, in general, creates pole of $u'$ triggering the
presence of singular solutions in some situations.  However, we
explicitly show that such singular solutions are physically
acceptable in a restricted domain of space labeled by $(x,t,u)$.
Substituting the expansion (\ref{proposed-expansion}) into
Equation~(\ref{2nd-order})  points to $m=2$ and so the
Lorentz-like expansion of $v(z)$ reads
\begin{equation}\label{proposed-v}
   v(z)=a_0+a_1F+a_2F^2+\frac{a_{-1}}{F}+\frac{a_{-2}}{F^2}\, .
\end{equation}

We next compute the expansion parameters $a_i$ that needs
term-by-term balancing of the coefficients of each powers of $F$
to zero. Somewhat involved but straightforward algebra leads to
the following relations
\begin{eqnarray}\label{a-relations}
  F^{\pm 4} &:& a_{\pm 2}[12\wedge_{\pm}+(\alpha+\beta)a_{\pm 2}]=0,\quad
           \left [\wedge_+\equiv \varepsilon_1\varepsilon_3 \right ]\, ,\nonumber \\
  F^{\pm 3} &:& a_{\pm 1}[2\wedge_{\pm}+(\alpha+\beta)a_{\pm 2}]=0, \quad\left [ \wedge_-\equiv \varepsilon_2\right ]\, , \nonumber \\
  F^{\pm 2} &:& 8\ws (\varepsilon_1\varepsilon_2+\varepsilon_3)a_{\pm 2}+(\alpha+\beta)\ws
               (a^2_{\pm 1}+2a_0a_{\pm 2})\nonumber \\
            && \mbox{}+2(1-\ws)a_{\pm 2}=0\, ,  \\
  F^{\pm 1} &:& \ws(\varepsilon_1\varepsilon_2+\varepsilon_3)a_{\pm 1}+(\alpha+\beta)\ws (a_0a_{\pm 1}+a_{\mp 1}a_{\pm 2})\nonumber \\
            &&\mbox{}+(1-\ws)a_{\pm 1}=0\, , \nonumber\\
  F^0 &:& 4\ws(\varepsilon_2a_2+\varepsilon_1\varepsilon_3 a_{-2})+(\alpha+\beta)\ws  ( a_0^2 \nonumber\\
      && \mbox{} +2\sum_{i=1}^2a_ia_{-i})+2(1-\ws)a_0+c_5=0\, .\nonumber
\end{eqnarray}

 \begin{table}[t]
 \caption{\label{table1}The solutions of the non-linear system (\ref{a-relations}) for the expansion
parameters $a_j,j=0,1,2$ are given. In the second column,
$\omega=\varepsilon_1\varepsilon_2+\varepsilon_3$.}
\begin{center}
 \begin{tabular}{@{}cccc}\hline \hline
 Class & $(\alpha+\beta)a_0$ & $a_1$ & $(\alpha+\beta)a_2$ \\ \hline
 $1$ & $1-\ws^{-1}-4\omega$ & $0$ & $-12\varepsilon_1\varepsilon_3$\\
 $2$ & $1-\ws^{-1}-4\omega$ & $0$ & $0$\\
 $3$ & $1-\ws^{-1}-4\omega$ & $0$ & $-12\varepsilon_1\varepsilon_3$\\
 $4$ & Arbitrary & $0$ & $0$\\\hline \hline
 \end{tabular}
 \end{center}
 \end{table}

By exploiting (\ref{a-relations}) we can derive four classes of
solutions for $a_j$ and $c_5$ which are summarized in
Table~\ref{table1} and Table~\ref{table2}. From a previous work we
already know that twelve different choices exist (see Table~I in
\cite{as4}) for the zeros of $F$ in right-hand side of
(\ref{constraint}) that leads to different representation for
$F(z)$ in terms of Jacobian elliptic functions. Further for each
of them one gets four classes of travelling wave solutions of GSWW
equation (\ref{GSWW}) corresponding to the solutions furnished in
Table~\ref{table1} and Table~\ref{table2}. Our new method is a
generalized procedure and encompasses the previously known
solutions as special cases. In the next subsection we provide, as
illustrative cases, four classes of solutions for three particular
choices of integral parameters in Equation~(\ref{constraint}).

\subsection{Class 1-4 solutions for
particular selections \\ of integral parameters}\label{particular}

To get explicit forms of Class 1--IV solutions obtained in EEF
method, it remains to choose the integral parameters
$\varepsilon_j,j=1,2,3$ in the constraint
$F'^2=(1+\varepsilon_1F^2)(\varepsilon_2+\varepsilon_3F^2)$. At
first let us consider some degenerate selections (i.~e.\ taking a
pair of double zeros of $F'$) leading to hyperbolic, trigonometric
and linear solutions.
\begin{itemize}

\item{$\underline{\mbox{Algebro-hyperbolic Waves}}$}

The choice $(\varepsilon_1,\varepsilon_2,\varepsilon_3)=(-1,1,-1)$
gives $F(z)=\tanh z$ that generate the following solutions modulo
a constant
\begin{equation}\label{degenerate-hyperbolic}
   u=(a_0+a_2+a_{-2})z-a_2\tanh z-a_{-2}\coth z\, .
\end{equation}
The above solution reduces to linear form for Class~4 while
singular term disappears in Class~3 solutions.

\item{$\underline{\mbox{Algebro-trigonometric Waves}}$}

Choosing $(\varepsilon_1,\varepsilon_2,\varepsilon_3)=(1,1,1)$, we
get $F=\tan z$ which gives following solutions apart from an
inessential factor
\begin{equation}\label{degenerate-trigonometric}
    u=(a_0-a_2-a_{-2})z+a_2\tan z-a_{-2}\cot z\, .
\end{equation}
\end{itemize}

 \begin{table}[t]
 \caption{\label{table2}The integration constant $c_5$ and
 $a_{-1},a_{-2}$ satisfying (\ref{a-relations}) are provided where
  $(\mathcal{A}_j-\omega^2)/(\varepsilon_1\varepsilon_2\varepsilon_3)=12/(-4)^{j-1}$,
  for $j=1,2$ and $\mathcal{A}_3=\ws a_0(\alpha+\beta)+2(1-\ws)$.}
  \begin{center}
 \begin{tabular}{@{}cccc}\hline \hline
 Class &  $a_{-1}$ & $(\alpha+\beta)a_{-2}$ & $\ws(\alpha+\beta)c_5$ \\ \hline
 $1$ & $0$ & $-12\varepsilon_2$ & $-(1-\ws)^2+16\ws^2\mathcal{A}_1$ \\
 $2$ & $0$ & $-12\varepsilon_2$ & $-(1-\ws)^2+16\ws^2\mathcal{A}_2$ \\
 $3$ & $0$ & $0$ & $-(1-\ws)^2+16\ws^2\mathcal{A}_2$ \\
 $4$ & $0$ & $0$ & $\ws a_0\mathcal{A}_3(\alpha+\beta)$\\\hline \hline
 \end{tabular}
 \end{center}
 \end{table}


Note that such types of algebro-hyperbolic and
algebro-trigonometric solutions are new and of interest. By
choosing suitable value of wave speed $\ws$ the algebraic term can
be removed causing the reduction to known forms obtained in
\cite{elw}. Now using a canonical procedure we have already
obtained [ see (\ref{hyperbolic-reduction-1}) \&
(\ref{periodic-reduction-2}) of Sec.~\ref{classical}] such types
of solutions in the extreme limits of modulus parameter of
elliptic functions. This means that more general forms of
solutions can be generated from EEF method provided $F'$ has four
simple zeros.

We already mentioned that twelve different selections are possible
for the simple zeros of $F'$ in the constraint (\ref{constraint})
leading to closed analytic expressions for $F(z)$ in terms of
doubly periodic Jacobian elliptic functions. Here we provide
explicit forms of the wave-solutions for the three representative
selections, two of which produce singular wave-solutions of
Class~$1$ and $2$. Note that given non-linear parameters
$\alpha,\beta$ we fix the the integration constant $c_5$ of
Equation~(\ref{2nd-order}) for a particular value of wave-speed
$\ws$ through a relation as dictated by the last column of
Table~\ref{table2} and hence that will not appear in the final
expressions of the solutions.
\begin{figure}[ht]
\begin{center}
 \includegraphics[height=10cm,width=10cm]{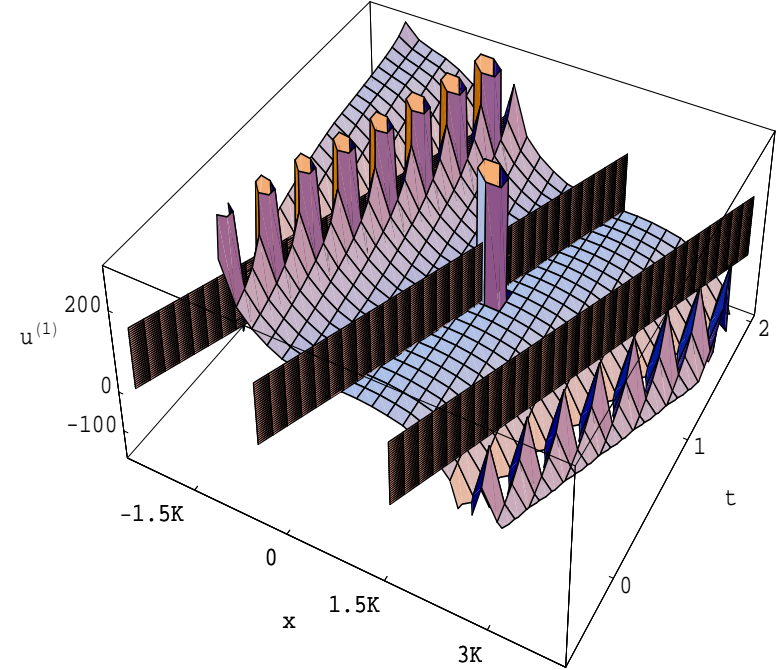}
 \end{center}
 \caption{\small Class~1 travelling wave for Selection~I. The primary inputs: $\alpha=0.5,\beta=0.1,k=0.7$ and the scaling parameters are
 $u_0=z_0=0$. The humps are the signature of singularity.} \label{I-1}
\end{figure}
\begin{itemize}
\item{\bf Selection I}

Let us choose the zeros of $F'$ as $\pm 1,\pm 1/k (0<k^2<1)$ which
 correspond to the selection of triplet
 $(\varepsilon_1,\varepsilon_2,\varepsilon_3)=(-1,1,-k^2)$ or $(-k^2,1,-1)$. Choosing $z_0$ such that
 $F(z_0)=0$, one then obtains (see Table~I of \cite{as4})
 \begin{equation}\label{sn-selection}
    F(z)=\sn (z-z_0)\, .
\end{equation}
Thus we are led to four classes of solutions of which the first
two ($j=1,2$) correspond to the singular solutions :
\begin{equation}\label{singular-1,2}
   \mbox{Class }j:\quad  v^{(j)}=a_0+a_2\sn^2(z-z_0)+a_{-2}\nss^2(z-z_0) \, ,
\end{equation}
where $a_i$s are obtained from 1st and 2nd rows of
Table~\ref{table1} \& Table~\ref{table2}. The corresponding
travelling wave solutions $u^{(j)}$ are obtained from the
integration of (\ref{singular-1,2}) and involves a singularity at
$z=z_0$ as is evident from the formula \cite{ste}
\begin{eqnarray}
     \int_0^{z-z_0} \left ( \nss^2\tau-\frac{1}{\tau^2}\right ) \textrm{d}
    \tau &=& z-z_0+\frac{1}{z-z_0}\nonumber\\
    &&\hspace{-4cm}\mbox{}-E(\vartheta_0,k)-\cn (z-z_0)\ds (z-z_0)\, .\label{formula}
\end{eqnarray}
In the expression (\ref{formula}) and also in the following
$\vartheta_{\ell}$ is defined by $\sin \vartheta_{\ell}=\sn
(z-z_{\ell},k)$ for $\ell=0,1$. In this context it is worth
mentioning that in the limit $k\rightarrow 1$, the travelling wave
reduces to $\coth$-type solution which also has a singularity at
$z=z_0$. This solution was derived from a different approach in
Ref.\ \cite{elw}. It is well known that such a singular solution
may serve as a model for physical phemenon of so-called
``hot-spots" \cite{smy,cla2,kud}.

In the following our aim would be to find regions where the
travelling wave solutions ($j=1,2$) corresponding to
(\ref{singular-1,2}) are regular. We use the following infinite
trigonometric series expansion \cite{ste} of $\nss (z-z_0,k)\equiv
1/\sn (z-z_0,k)$ :
\begin{eqnarray}
     \nss (z-z_0,k)&=&\frac{\pi}{2K}\textrm{cosec} \tau \nonumber \\
     && \hspace{-2cm}\mbox{}-\frac{2\pi}{K}\sum_{n=0}^{\infty}\frac{q^{2n+1}}{1-q^{2n+1}}\sin
    [(2n+1)\tau]\, ,\label{ns-expansion}
\end{eqnarray}
where $q=\exp [-\pi K'/K]$ is known as nome of the elliptic
functions and  $\tau=\pi(z-z_0)/2K$, $K'(k)=K(k')$, the quantities
$K(k)$ and $k$ have already been defined in Sec.~\ref{classical}.
Denoting the singular term in the travelling wave solutions by the
symbol $S_0$, we see that as $z\rightarrow z_0$, the asymptotic
expression of its 1st term will be
\begin{equation}\label{asymptotic1}
    S_0(z\rightarrow z_0)=\frac{\pi}{2K}\textrm{cosec}\tau
\end{equation}
We will now use the well known expansion for $\textrm{cosec}\tau$
convergent in the region $|\tau|<\pi$
\begin{equation}\label{asymptotic2}
  \textrm{cosec}\tau=\frac{1}{\tau}+\sum_{n=1}^{\infty}\frac{(-1)^{n-1}2(2^{2n-1}-1)B_{2n}}{(2n)!}\tau^{2n-1}\, ,
\end{equation}
where $\{ B_n\}$ are Bernoulli numbers whose first few members are
$B_0=1,B_1=-1/2$, $B_2=1/6,B_4=-1/30$ etc.

Combining (\ref{asymptotic1}) and (\ref{asymptotic2}), it is
straightforward to conclude that near $z\sim z_0$, the travelling
wave solution behaves like $(z-z_0)^{-1}$ i.~e. the point $z=z_0$
is a simple pole.
Note that the region of convergence of the series
(\ref{asymptotic2}) for $\textrm{cosec}\tau$ is
\begin{equation}\label{convergence-region}
    |\tau|<\pi\Rightarrow |z-z_0|<2K
\end{equation}
The above equation speaks of two parallel planes
\begin{equation}\label{parallel-planes}
    x-ct-z_0\pm2K=0\, ,
\end{equation}
which are normal to wave surface. These two planes are shown in
Fig.~\ref{I-1} [see two extreme strips orthogonal to surface
$u=u(x,t)$]. It is thus quite expected that wave surface will be
irregular along these two planes. This fact can be appreciated
from the humps along the strips in Fig.~\ref{I-1} showing sudden
rise of the value $u=u(x-ct)$. Further there must be a singularity
at $z=z_0$. From elementary analysis we know that in $t$-$x$ plane
the point $z=x-ct$ approaches point $z_0$ along any curve. Hence
there will be a circular disc in the plane $z=z_0$ centred at
$z_0$ and of arbitrary small radius where solution will show
irregularity. This is reflected by a single localized hump in
Fig.~\ref{I-1} on the plane lying parallel and midway between two
extreme planes.

To use such a solution we thus need an exact knowledge about the
position of parallel humps. These may be exactly computed by the
solid angles $\theta_x=\cos^{-1}|1/\sqrt{1+\ws^2}|$ and
$\theta_t=\cos^{-1}|\ws/\sqrt{1+\ws^2}|$. These two angles in turn
depend on the parameter $\ws$ which determines the wave speed.
Hence  very fast and very slow waves have different regions of
humps. The core singularity at $z=z_0$ will however be present in
same region for all types of waves which may be removed by making
a hole of arbitrary small radius on the plane $z=z_0$ around point
$z_0$. This concludes our detailed analysis about singularity of
Class~1 and 2 solutions for Selection~I which hopefully provides a
sufficient basis of the solution regarding its potential use.

The other two classes correspond to the non-singular solutions
\begin{equation}\label{nonsingular-1,2}
    \mbox{Class }j: \quad v^{(j)}=a_0+a_2F^2 \, , \qquad j=3,4\, .
\end{equation}
The resulting travelling wave solutions are of the same form as
the solutions (\ref{final-sol-1}) and (\ref{linear-reduction-2})
obtained in Sec~\ref{classical}:
\begin{equation}\label{nonsingular-wave-1,2}
     u^{(j)} = u_0+(a_0+\frac{a_2}{k^2})(z-z_0)-\frac{a_2}{k^2}E(\vartheta_0,k)\, ,
\end{equation}
where $a_i$s are to be computed from last two rows of
Table~\ref{table1} and Table~\ref{table2}. Note that Class~$4$
solution is linear in $x$ and $t$, since $a_{\pm 1}=a_{\pm 2}=0$
(see the last rows of the tables).

\item{\bf Selection II}

 Let us now choose pairs of purely real and of purely imaginary roots of $F'$ respectively
 as $\pm 1/k' (0<k'^2=1-k^2<1)$ and $\pm  i/k$ which come from the selection $(\varepsilon_1,\varepsilon_2,\varepsilon_3)=(-k'^2,1,k^2)$
 or  $(k^2,1,-k'^2)$. One then obtains the following representation of $F(z)$:
\begin{equation}\label{sd-selection}
    F(z)=\sd (z-z_0,k)\, ,
\end{equation}
leading to a new solution of $u(z)$. The explicit expressions for
Class~1 and Class~2 are given by
\begin{eqnarray}
  u^{(j)}&=& u_0+\varrho_j (z-z_0)-(3-j)a_{-2}E(\vartheta_0,k)\nonumber \\
  && \hspace{-1.3cm}\mbox{}+\!(2\!-\!j)k^2a_{-2}\sn (z\!-\!z_0,k)\cd (z\!-\!z_0,k)\!+\!S_0\label{singular-wave-3}
\end{eqnarray}
for $j=1,2$, where we have abbreviated $\varrho_j$ as
$\varrho_j=a_0+(3-j)k'^2a_{-2}$. The non-singular solutions
corresponding to Class~3 and 4 solutions read
\begin{eqnarray}\label{singular-wave-4}
    u^{(j)} &=& u_0+\left ( a_0-\frac{a_2}{k^2}\right )(z-z_0)+\frac{a_2}{k^2k'^2}E(\vartheta_0,k)\nonumber \\
  && \hspace{-1cm}\mbox{}-\frac{a_2}{k'^2}\sn (z-z_0,k)\cd (z-z_0,k)\, ,\: j=3,4\, .
\end{eqnarray}

The main feature of Selection~II is that the solutions coming from
that  are nearly periodic in the whole domain due to the
additional sn and cd-term. In contrast the solutions generated
from Selection~I are only quasi-periodic, the periodic behaviour
is observed for singular solutions which are prominent near $z\sim
z_0$.

\begin{figure}[t]
\begin{center}
 \includegraphics[height=10cm,width=10cm]{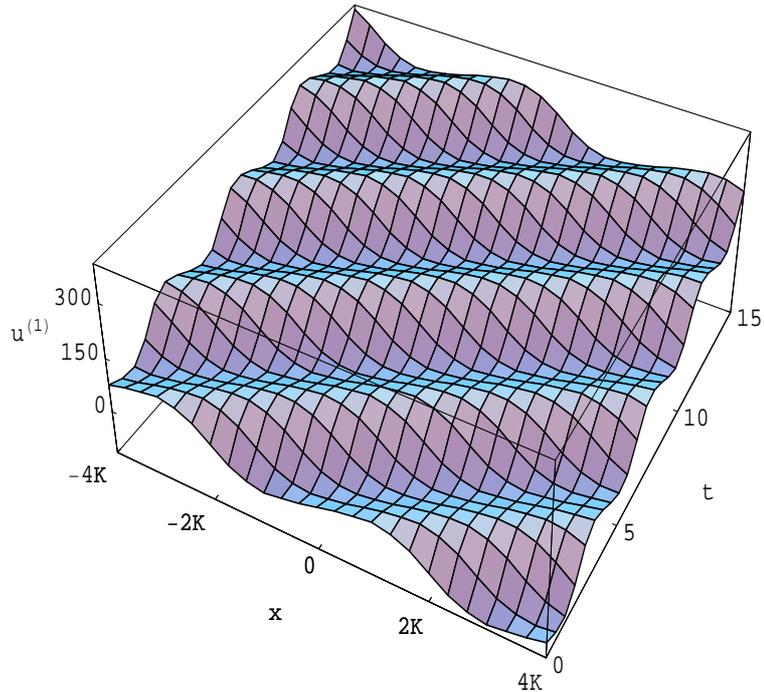}
 \end{center}
  \caption{\small Class~1 travelling wave for Selection~III for the same primary inputs
 and scalings are $u_1=z_1=0$. The smoothness of the wave-surface shows the solution is non-singular.} \label{III-3}
\end{figure}

\item{\bf Selection III}

We just saw that the Class~$1$ and $2$ solutions generate singular
solutions owing to the generating functional $F(z(x,t))$
possessing either a zero or a pole in the region of validity. It
is indeed possible to generate non-singular solutions for these
classes, which share similar qualitative behaviour with the
previous counterparts, if $F$ can be chosen to have neither zero
nor a pole in the finite part. Below we provide such an example,
which are the only one among the set of elliptic functions.

We choose the zeros of $F'$ as $\pm 1,\pm 1/k'$ corresponding to
the selection
$(\varepsilon_1,\varepsilon_2,\varepsilon_3)=(-k'^2,-1,1)$. Then
we have the following representation of $F(z)$ (see Table~I of
\cite{as4})
\begin{equation}\label{dn-selection}
    F(z)=\nd (z-z_1,k)\, ,
\end{equation}
where $z_1$ is fixed as $F(z_1)=1$. It is now trivial to check
that the generated solutions for each of the four classes remain
non-singular
\begin{eqnarray}
  u^{(j)} &=& u_1+a_0(z-z_1)+(3-j)a_{-2}E(\vartheta_1,k) \nonumber \\
  && \hspace{-0.9cm}\mbox{}-(2\!-j)k^2a_{-2}\sn (z\!-\!z_1,k)\cd (z\!-\!z_1,k)\, ;\label{nonsingular-wave-5}\\
  u^{(j)} &=& u_1+a_0(z-z_1)+a_2E(\vartheta_1,k)/k'^2\, , \nonumber \\
  && \hspace{-0.5cm}\mbox{}-k^2a_2\sn (z-z_1,k)\cd (z-z_1,k)/k'^2\, ,\label{nonsingular-wave-6}
\end{eqnarray}
where $j=1,2$ for  Equation~(\ref{nonsingular-wave-5}) and $j=3,4$
for (\ref{nonsingular-wave-6}). Fig.~\ref{III-3} depicts a Class~1
wave guided by (\ref{nonsingular-wave-5}).
\end{itemize}

Other possible selections from Table~I of \cite{as4} will generate
many such new elliptic travelling wave solutions of GSWW equation.
Let us mention that the degenerate selections leading to
hyperbolic and trigonometric type waves can be obtained from
 elliptic solutions  in
 $k\rightarrow 0,1$ limits.
 \section{Summary}\label{conclusion}

 In this article we proposed an extended method to generate a rich class of doubly-periodic elliptic
 travelling wave solutions of the GSWW equation. A systematic classification is given for the solutions
 to exhaustively utilize the strength of the proposed method.  We also discussed a canonical
 procedure for generating travelling wave solutions. The problem with such solutions is that these require the
 pre-knowledge of zeros of second derivative of the solutions. The proposed  EEF method
 removes this difficulty by fixing the zeros of the first derivative of the solutions and leads to a wide range of travelling
 waves which include the kink type solitary wave  \cite{miu}, sinusoidal type periodic solution \cite{cla} and a rational
 solution (\cite{dra}). Class~1 to 4 classify the various types of solutions whose explicit forms are
 noted for three representative selections of integral parameters
 defining the generating functional. Classes 1 and 2 produce singular solutions which has
 important application in modelling the physical curcumstances of formation of
 hot-spots \cite{smy,cla2,kud}. We have exactly computed the positions of such
 blow-up of solutions and graphically illustrated in the Fig.~\ref{I-1}. To the best of our knowledge
 such an analysis about the region of blow-up for singular elliptic solutions has not been
 done previously in the literature.

 \end{document}